\def\e{{\rm e}}
\def\del{\partial}

\def\abs#1{{\left|{#1}\right|}}

\newcommand{\NPB}[3]{Nucl. Phys. {\bf B{#1}} (19{#2}){#3}} 
\newcommand{\PLB}[3]{Phys. Lett. {\bf B{#1}} (19{#2}){#3}} 
 
\newcommand{\ANN}[3]{Ann. Phys. {\bf {#1}} (19{#2}){#3}} 
\documentstyle[12pt]{article}
\begin{document}
\title{Supersymmetry Breaking through
Boundary Conditions Associated with the $U(1)_{R}$} \author{Kazunori 
Takenaga $^{}$\thanks {email: takenaga@oct.phys.kobe-u.ac.jp,~~JSPS 
Research Fellow}\\ \it {Department of Physics, Kobe University,}\\ 
{\it Rokkodai, Nada, Kobe 657, Japan}
}
\date{} 
\maketitle
\baselineskip=18pt
\begin{abstract}
The effects of boundary 
conditions of the fields for the compactified space directions
on the supersymmetric theories are discussed. The boundary conditions 
can be taken
to be periodic up to the
degrees of freedom of localized $U(1)_{R}$ transformations. The 
boundary condition
breaks the supersymmery to yield universal soft supersymmetry 
breaking terms. The $4$-dimensional supersymmetric QED with one 
flavour
and the pure supersymmetric QCD are studied as toy models when one of 
the space coordinates is compactified on $S^1$. \end{abstract}
\addtolength{\parindent}{2pt}
\newpage
\par
In this paper we investigate the effects of boundary conditions 
of fields on supersymmetry breaking when one of the space 
coordinates is compactified on $S^{1}$. One does not know, {\it a 
priori}, what boundary conditions should be imposed on fields for the 
$S^{1}$-direction. We shall consider general boundary conditions 
except for the periodic one.
This is very contrary to the case of the finite temperature field 
theory, in which the boundary conditions for the euclidean time 
direction is
determined definitely by the quantum statistics of particles. \par
As general and possible boundary conditions, one can require that the 
fields return to their original values up to phases 
\footnote{We ignore phases for the 
fermion fields associated with the continuous spin structure of the 
manifold in this paper.}
proportional to their charges of global symmetry 
transformations\cite{ahosotani}
when the fields travel along the $S^{1}$ direction.
The global symmetry transformations must be symmetry of the theories .
The lagrangian is still single-valued even if the fields have such the boundary 
conditions.
The boundary conditions are periodic for the $S^{1}$
direction up to the degrees of freedom of local transformations. In 
other words, the fields with the boundary conditions are mutually 
related with 
the fields with the periodic boundary condition 
by transformations which are obtained 
by localizing the aforementioned global symmetries and 
whose parameters
depend linearly only on the compactified coordinate \cite{schwartz}. 
\par
Having the boundary conditions mentioned above, the translational 
invariance may be broken for a certain boundary condition. The 
translational invariance, however, crucial for the supersymmetric 
invariance.
In supersymmetric theories the variations of action under the 
supersymmetric transformations vanish up to total derivative terms. 
If the translational invariance is broken for some compactified 
directions due to the boundary conditions, the total
derivative terms do not vanish and
remain as surface terms.
The supersymmetry is explicitly broken in this case.
Thus, the boundary conditions may break the supersymmetry. \par
One may think that we can use the boundary conditions associated with 
the global gauge symmetry in supersymmetric gauge theories. The 
boundary conditions, however, do not break the translational 
invariance.
One can always redefine the fields so as to satisfy 
the periodic boundary condition by 
local gauge transformations whose 
parameters depend linearly on the compactified coordinate. Since 
the local gauge
symmetry is the symmetry of the theory, the effects of non-trivial 
phases disappear and the supersymmetry is not broken. \par
In order to have possibilities to break the supersymmetry by boundary 
conditions, their charges of global symmetry transformations 
must be different
between the bosons and the fermions in a supermultiplet. The 
$U(1)_{R}$ symmetry, which is global symmetry of the theory, is the 
candidate in supersymmetric theories. The total derivative terms do 
not return to their original values
after the translation along the $S^{1}$ direction because of the 
charge differences between the bosons and the fermions in a 
supermultiplet. And they remain as the surface terms. The 
translational invariance
is, thus, broken for the $S^{1}$ direction in this case. Therefore, 
the supersymmetry is explicitly broken by the boundary condition.
\par
All the effects of the boundary condition associated with the $U(1)_{R}$ 
symmetry turn
out to appear as universal soft supersymmetry breaking terms.
One can redefine the fields so as to satisfy the periodic 
boundary condition by localized $U(1)_{R}$ transformation. Because the 
$U(1)_{R}$ symmetry is not the local symmetry, the transformation is 
not
respected as a symmetry of the theory by the terms including the 
ordinary
derivative, $\del_{\mu}$. The supersymmetry breaking
terms are generated only through the derivative.
As the consequence, such generated terms have couplings with mass dimensions 
and are the soft supersymmetry breaking terms since a derivative has mass 
dimension one. Moreover, the derivatives are the same for all 
flavours, so that the generated terms are common to all flavours 
which may be needed to avoid the FCNC. It 
should be stressed that once one has the boundary condition, such 
desirable supersymmetry breaking terms are automatically incorporated 
into theory with an unique parameter, $U(1)_{R}$ "couplings". \par
Now, let us see how the boundary condition associated with the 
$U(1)_{R}$ symmetry actually breaks the supersymmetry and how the 
soft supersymmetry breaking terms appear. We shall study the 
$4$-dimensional supersymmetric QED (SQED)
with one flavour and the pure supersymmetric QCD (SQCD) as toy 
models when one of the 
space coordinates, say, $x^3\equiv y$ is compactified on $S^{1}$ 
whose length is $L$.
\par
First, let us study the SQED with no
flavours. The lagrangian of the $4$-dimensional supersymmetric QED is 
constructed by the $F$-term of $W_{A}(\theta)^2$, where the chiral 
spinor
superfield, $W_{A}(\theta)$ contains the vector boson, $V_{\mu}$ 
(photon), a two-component Weyl fermion, $\lambda_{A}~(A=1, 2)$ 
(gaugino) and the auxiliary 
field, $D$ in the Wess-Zumino gauge \cite{wessb}. The $\theta$ is the 
superspace coordinates. Under the supersymmetric transformation, 
$\delta_{\xi}$ in uncompactified $4$-dimension, the on-shell 
lagrangian, ${\cal L}_{SQED}$ varies as
$\delta_{\xi} {\cal L}_{SQED}=\del_{\mu} X^{\mu}$, where $X^{\mu}$ is 
calculated as
\begin{equation}
X^{\mu}(\xi, \lambda,V_{\mu})
=-\xi\sigma_{\nu}{\bar\lambda}V^{\mu\nu} +{i\over 2}
\xi{\sigma^{\rho\sigma}}
\sigma^{\mu}{\bar\lambda}V_{\rho\sigma}+h.c.\quad . \label{eq:total}
\end{equation}
Here the $\xi$ is the supersymmetric
transformation parameters of
two-component constant Weyl spinors
and $V_{\rho\sigma}$ is the field strength for the photon. \par
The $U(1)_{R}$ symmetry is defined by
$W_{A}(\theta) \rightarrow
{\rm e}^{i
\beta}W_{A}({\rm e}^{-i \beta}\theta).
$
It is obvious that $\lambda$ has a $U(1)_{R}$ charge and 
$V_{\mu}$ and $D$ are neutral under the symmetry. Using the symmetry, 
we define the boundary conditions of the fields for the $S^{1}$ 
direction as follows; \begin{equation}
V_{\mu}(x^i, y+L)=V_{\mu}(x^i, y),\quad
\lambda(x^i, y+L)={\rm e}^{i\beta}\lambda(x^i, y). 
\label{eq:condition}
\end{equation}
The lagrangian we consider is still singled-valued even if we take the 
boundary condition, (\ref{eq:condition}). 
The superfield, $W_{A}(\theta)$ itself does 
not return to its original values after a translation along the 
$S^{1}$-direction, but the $F$-terms of $W(\theta)^2$ by which the 
supersymmetric lagrangian is constructed contain only the bilinear 
form, $\lambda\sigma^{\mu}\del_{\mu}{\bar\lambda}$. Therefore, the 
non-trivial phase, ${\rm e}^{i\beta}$ disappears when the lagrangian 
travels along the $S^{1}$-direction.
Having the boundary condition
(\ref{eq:condition}), the surface term,
\begin{equation}
\delta_{\xi}S_{SQED}=
\bigl(\int d^{3}{\bf x}~X^3
(\xi, V_{\mu}, \lambda)(x^{i}, y)\bigr)\mid{_{S^{1}}} 
\label{eq:surface} \end{equation}
does not vanish because there is a difference between $X^{3}(x^{i}, 
y+L)$ and
$X^{3}(x^{i}, y)$ due to the non-trivial phases of 
(\ref{eq:condition}).
Here the $X^3(\xi, V_{\mu}, \lambda)$ is the third space-component of the 
total derivative terms, $X^{\mu}$. Note that $\xi$ obeys the 
periodic boundary condition. The $x^i (i=0, 1, 2)$ stands for 
uncompactified coordinates. The translational invariance for the
$S^{1}$ direction is broken by the boundary condition. \par
When we expand the fields in the Fourier series for $S^1$ direction, 
$$
V_{\mu}(x^{i}, y)={1\over{\sqrt L}}\sum_{n=-\infty}^{+\infty} 
V_{\mu}^{(n)}(x^{i})\e^{{{2\pi i}\over L}ny},\quad \lambda(x^{i}, 
y)={1\over{\sqrt L}}\sum_{n=-\infty}^{+\infty} 
\lambda^{(n)}(x^{i})\e^{{{2\pi i}\over L}(n+{{\beta} \over 
{2\pi}})y}, $$
we see that $\lambda(x^{i}, y)$ is always redefined so as to satisfy 
the periodic boundary condition by the 
local transformation whose parameters depend linearly only 
on the compactified coordinate $y$;
\begin{equation}
\lambda (x^i, y)=U_{R}(y){\tilde{\lambda}}(x^i, y) \quad
{\rm with}\qquad U_{R}(y)\equiv {\rm e}^{i{\beta \over L}y}, 
\label{eq:trans}
\end{equation}
where $\tilde\lambda(x^i, y)$ satisfies
${\tilde{\lambda}}(x^i, y+L)={\tilde{\lambda}}(x^i, y)$. The relation 
(\ref{eq:trans}) is the same with those of the coordinate dependent 
compactifications used in the supergravity \cite{schwartz} and 
superstring \cite{bachas}. But we find that the effects of the 
boundary condition to supersymmetric theories are remarkable which 
will be discussed below.
\par
Suppose for a moment that the $U(1)_{R}$ is a local symmetry of the
theory. The gauge field, $V^R_{\mu}$ is necessarily introduced into the theory.
In this case one can always redefine the field, $\lambda (x^i, y)$ so 
as to satisfy the periodic boundary condition by the local transformation, 
$U_{R}^{\dagger}(y)$ without contradicting the local symmetry of the 
theory. The local transformation, $U_{R}^{\dagger}(y)$, which is now 
a part of the local
gauge transformation, shifts only the gauge field, $V^{R}_{3}$ by the 
constant, $\beta/eL$. The shift is compensated by the gauge 
transformation of the original gauge field, $V^R_{\mu}$. Here the $e$ 
is a gauge coupling constant. 
One can say that
the theory written in terms of $(V_{\mu}, \lambda)$ is equivalent to 
the theory written in terms of $(V_{\mu}, \tilde\lambda)$ thanks to the 
local gauge invariance of the theory.
The boundary condition does not break the translational invariance, 
nor the supersymmetric invariance.
\par
But the $U(1)_{R}$ symmetry is not actually a local symmetry of the 
theory.
It is impossible to make the boundary condition periodic by the local 
transformation, $U_{R}(y)$ keeping the equivalence of the theories with 
the two different boundary conditions.
Once we impose the boundary condition,
(\ref{eq:condition}), the translational invariance is broken for the 
$S^{1}$ direction and the supersymmetry is broken explicitly by
the surface terms, (\ref{eq:surface}). How do the effects 
of the breaking manifestly appear in the lagrangian? 
The supersymmetry transformations are no longer defined
in terms of $V_{\mu}$ and $\lambda$
because they do not satisfy the same boundary condition.
In order to understand the supersymmetry breaking at the lagrangian 
level, one needs to have the supersymetry transformations.
We can define the supersymmetry transformations between $V_{\mu}$ 
and the redefined field, $\tilde\lambda$ by forming $V_{\mu}(x^i, y)$ and 
${\tilde\lambda}(x^i, y)$ into a supermultiplet.
The two fields satisfy the same boundary condition, say, periodic 
boundary condition. The effects of the supersymmetry breaking 
can be understood as a 
difference from the periodic boundary condition.   
One can not manifest the effects of the supersymmetry breaking 
through the 
boundary condition until the supersymmetry transformation 
is defined in terms of the fields satisfied by the same 
boundary condition between the bosons and the fermions in a 
supermultiplet. \par
By redefining $\lambda$ by (\ref{eq:trans}) into $\tilde\lambda$ , the 
variation of the lagrangian under the modified 
supersymmetric transformations becomes as follows;
\begin{equation}
\tilde{\delta_{\xi}}{\cal L}_{SQED}
=\del_{\mu}X^{\mu}(\xi, V_{\mu},
\tilde{\lambda})+U_{R}\del_{\mu}
U_{R}^{\dagger}[i\xi\sigma^{\rho\sigma}\sigma^{\mu}{\tilde{\bar\lambda}} 
V_{\rho\sigma}+h.c.], \label{eq:periodic} \end{equation}
where $\tilde{\delta_{\xi}}$ defines the modified supersymmetric 
transformations in terms of $V_{\mu}$ and $\tilde\lambda$.
The first term in (\ref{eq:periodic}) does not generate surface terms 
because all the fields satisfy the periodic boundary condition. The 
boundary condition associated with the $U(1)_{R}$ symmetry breaks the 
supersymmetry
explicitly as shown in the second term in (\ref{eq:periodic}). As we 
expected, the breaking of the supersymmetry is entirely due to the 
locality
of the $U(1)_{R}$ transformation, {\it i.e.} $U_{R}\del_{\mu} 
U_{R}^{\dagger}$. If the $U(1)_{R}$ is a local symmetry of the 
theory, the second term in (\ref{eq:periodic}) is
absorbed into the gauge field associated with the gauged $U(1)_{R}$ 
symmetry.
\par
Let us discuss the supersymmetry breaking terms. In terms of redefined 
field, $\tilde\lambda$, the ${\cal L}_{SQED}$
can be rewritten as
\begin{equation}
{\cal L}_{SQED}(V_{\mu},\lambda)
={\cal L}_{SQED}(V_{\mu}, \tilde\lambda) +{\cal L}_{SQED}^{soft},
\label{eq: soft}
\end{equation}
where ${\cal L}_{SQED}(V_{\mu},
\tilde\lambda)$ is the same with the original lagrangian except that 
all the fields satisfy the periodic boundary condition, and ${\cal 
L}_{SQED}^{soft}$ is obtained as
$$
{\cal L}_{SQED}^{soft}=-i[U_{R}(y)\del_{\mu}U_{R}^{\dagger}(y)] 
{\tilde\lambda}\sigma^{\mu}{\tilde{\bar\lambda}} =-{\beta\over 
L}{\tilde\lambda}\sigma^3{\tilde {\bar\lambda}} =-{\beta\over 
L}(\bar{\psi_{1}}\psi_{1}+\bar{\psi_{2}}\psi_{2}), $$
where in the last equality we have used the Majorana spinors in the 
$3$-dimension defined by ${\tilde\lambda}_{M}^T=(\psi_{1}, i\psi_{2})^T$. 
The $\lambda_{M}$ is the $4$-component Majorana spinors in the 
$4$-dimension constructed by ${\tilde\lambda}_{M}^T\equiv 
({\tilde\lambda}_{A}, {\tilde\lambda}^{\dot A})^T$.
We note that the periodicity in $\beta$ is recovered by the redefinition of the 
gaugino field, $\lambda\rightarrow {\rm e}^{2{\pi}iny/L}\lambda$. 
The ${\cal L}_{SQED}^{soft}$ is generated through the derivative, 
$\del_{\mu}$ in the kinetic term 
for the gaugino, which does not respect the invariance of the theory 
under the local transformation, $U_{R}(y)$. Therefore, the kinetic 
term is the only source for the supersymmetry breaking. As the 
remarkable consequence, the supersymmetry breaking terms generated in 
this mechanism are common to all flavours. Moreover, the breaking is 
always so-called soft breaking. This is because the derivative 
has mass dimension one, so that the 
couplings generated through the derivative are always dimensional 
couplings.
Note that the second term in (\ref{eq:periodic}) coincides with the 
modified supersymmetric transformations of ${\cal L}_{SQED}^{soft}$. 
If the $U(1)_{R}$ is a local symmetry of the theory, 
the term is absorbed into the gauge field associated with the gauged 
$U(1)_{R}$ symmetry.
\par
Let us briefly discuss effects of the boundary condition arising from a flavour 
multiplet on the supersymmetry breaking. The theory we consider is the  
SQED with one massive flavour. We introduce two massive chiral superfields, 
$\Phi_{i}~(i=1, 2)$ and assign the gauge charge $+e$ for the 
$\Phi_{1}$ and $-e$ for the $\Phi_{2}$. The gauge invariant 
superpotential is given by $W=m\Phi_{1}\Phi_{2}$. For each chiral 
superfield, the $\Phi_{i}$ contains a complex scalar, 
$A_{i}$(selectron), a two-component Weyl spinors, $\chi_{i}$(electron) and 
the auxiliary field, $F_{i}$. 
As discussed before, the supersymmetry is not broken by the 
boundary condition associated with the global gauge symmetry, 
which corresponds to the conservation of the lepton number. 
The $U(1)_{R}$ symmetry is only operational for its breaking. We can define the 
$U(1)_{R}$ symmetry by $\Phi_{1}(\theta)\rightarrow
{\rm e}^{i\beta}\Phi_{1}({\rm e}^{-i\beta}\theta)$ and 
$\Phi_{2}(\theta)\rightarrow
{\rm e}^{i\beta}\Phi_{2}({\rm e}^{-i\beta}\theta)$. 
It is obvious 
that the fermion field, $\chi_{i}$ does not carry 
the $U(1)_{R}$ charge and the complex scalar field, $A_{i}$ and the 
auxiliary field, $F_{i}$ carry the $U(1)_{R}$ charge. \par
The field, $A_{i}$ is always redefined so as to satisfy the periodic 
boundary condition by the local transformation,
$A_{1}(x^i, y)=U_{R}(y){\tilde A_{1}}(x^i, y),~ A_{2}(x^i, 
y)=U_{R}(y){\tilde A_{2}}(x^i, y)$, where $U_{R}(y)\equiv 
{\rm e}^{i\beta/L}$.
The derivative in the kinetic term for $A_{i}$ does not respect 
the invariance under the local transformation, $U_{R}(y)$, so that 
the soft supersymmetry breaking term are generated from there. Hence, 
the result is given by
$$
{\cal L}_{SQED}^{flavour}=({\beta \over L})^2 
({\abs {\tilde{A_1}}}^2+{\abs {\tilde{A_2}}}^2)
+{{2e\beta}\over L}({\abs {\tilde{A_1}}}^2-{\abs {\tilde{A_2}}}^2)\phi, 
$$
where
the third space-component of the gauge field, $V_{3}$ is denoted by $\phi$. 
We find that the soft supersymmetry breaking terms are the scalar 
mass terms and the trilinear scalar couplings which are common to all 
flavours. The supersymmetry breaking terms depend on an 
unique parameter, $\beta$ and the gauge coupling, $e$. \par
Finally, let us discuss the SQCD with the gauge group $SU(N_{c})$ and 
no flavours. The discussions are almost the same with the case for 
the SQED.
One can take the boundary conditions of the fields as follows
\cite{ahosotani};
\begin{equation}
A_{\mu}(x^i, y+L)=U_{g}A_{\mu}(x^i, y)U_{g}^{\dagger},\quad 
\lambda(x^i, y+L)={\rm e}^{i\beta}U_{g}\lambda(x^i, y) 
U_{g}^{\dagger},\label{eq:bcondition}
\end{equation}
where $U_{g}\in SU(N_{c})$. 
The non-trivial phase ${\rm e}^{i\beta}$ is associated with the 
$U(1)_{R}$ symmetry.
As for the factor, $U_{g}$ associated with the global gauge 
symmetry, it is shown to give no physical effects at least 
classically, {\it i.e.}, the fields can always be taken to be 
periodic by utilizing the freedom of the local gauge transformation 
whose parameters depend linearly only on $y$.
The factor, however, is actually related with the non-integrable phases of the 
gauge field along the $S^{1}$ direction whose effects are essential
at the quantum level for studying the local gauge symmetry 
breaking of the theory \cite{bhosotani}.
Let us see how the softly supersymmetry breaking terms are generated 
by the boundary condition.
In terms of the redefined fields, the lagrangian can be rewritten as $$
{\cal L}_{SQCD}(V_{\mu}, \lambda)
={\cal L}_{SQCD}(V_{\mu}, \tilde\lambda) +{\cal L}_{SQCD}^{soft},
$$
where ${\cal L}_{SQCD}(V_{\mu}, \tilde\lambda)$ is the same with the 
original lagrangian
except that $\lambda$ is replaced by $\tilde\lambda$ which satisfies the 
periodic boundary condition. The ${\cal L}_{SQCD}^{soft}$ is obtained 
as $$
{\cal L}_{SQCD}^{soft}=-2i[U_{R}(y)\del_{\mu}U_{R}^{\dagger}(y)] {\rm 
tr}[{\tilde\lambda}\sigma^{\mu}
{\tilde {\bar\lambda}}]=-{2\beta\over L}{\rm tr} [{\tilde 
\lambda}\sigma^3{\tilde {\bar\lambda}}] =-{2\beta\over L}{\rm 
tr}(\bar{\psi_{1}}\psi_{1}+\bar{\psi_{2}}\psi_{2}), $$
where we have used the definition of the $3$-dimensional Majorana 
spinors as before.
Here again, one can conclude that the soft supersymmetry breaking 
terms are generated through the kinetic term of the gaugino. \par
We have discussed that the supersymmery can be broken explicitly by 
the boundary conditions associated with the $U(1)_{R}$ symmetry.
The $U(1)_{R}$ charges are different between the bosons and fermions 
in a supermultiplet. Then, the translational invariance for the 
$S^{1}$ direction is broken by the boundary condition, so that the 
surface terms of the total derivative remain. 
One obtains all the effects of the boundary condition as the soft 
supersymmetry breaking terms in the lagrangian. 
This can be shown explicitly by 
redefining the fields so as to satisfy the periodic boundary 
condition by the local transformation, (\ref{eq:trans}).
The effects are always soft supersymmetry breaking because they are 
generated only through the derivative, $\del_{\mu}$, which has mass 
dimension one, in the kinetic term for the gaugino and selectron, where 
the local transformation, $U_{R}(y)$ is not respected as the symmetry of the 
theory. Remarkable feature is that the soft supersymmetry breaking terms 
do not have many arbitrary parameters, as usually discussed, but they 
depend on an unique parameter, $\beta$ and the gauge coupling. 
These soft supersymmetry breaking terms 
are common to all flavours because the terms are generated from the 
same origin, that is, the derivative in the kinetic term.
It is desirable to avoid the FCNC. 
It should be stressed that these soft supersymmetry 
breaking terms are automatically incorporated into the theories by 
the boundary conditions associated with the $U(1)_{R}$ symmetry.\par 
The boundary conditions, $U_{g}$ in (\ref{eq:bcondition}) are related 
with the non-integrable phases of the gauge fields along the $S^{1}$ 
direction. One can study how the local gauge symmetry of the theory 
with the soft supersymmetry breaking terms discussed here is 
spontaneously broken through the dynamics of the non-integrable 
phases \cite{bhosotani}. We believe that there are new possibilities 
for building models with soft supersymmery breaking terms in more 
realistic higher dimensional super Yang-Mills theories \cite{brink}.
These issues are under investigations and will appear soon. \par
\begin{center}
{\bf Acknowledgment}
\end{center}
\vspace{12pt}
The author would like to thank Professor C. S. Lim for fruitful 
discussions, encouraging me and reading
this manuscript. This work was supported by Grant-in-Aid for 
Scientific Research Fellowship, No.5106. 


\begin{thebibliography}{99}
\bibitem{ahosotani} Y. Hosotani, \ANN{190}{89}{233}. 
\bibitem{schwartz} J. Scherk and J.H. Schwartz, \PLB{82}{79}{60}. 
\bibitem{wessb} See, for example, J. Wess and J. Barger,~ 
Supersymmetry and Supergravity, Princeton University Press. 
\bibitem{bachas} C. Bachas, hep-th. 9503030, references there in. 
\bibitem{bhosotani} Y. Hosotani, \PLB{126}{83}{309}. \bibitem{brink} 
L. Brink, J. H. Schwartz and J. Scherk, \NPB{121}{77}{77}. 
\end{thebibliography}
\end{document}